\documentclass[preprintnumbers,aps,prl,twocolumn,showpacs,floatfix,nofootinbib,superscriptaddress,groupedaddress]{revtex4}  

\pdfoutput=1 

\usepackage[tbtags]{amsmath}  
\usepackage{amssymb}          
\usepackage{bm}               
\usepackage{graphicx}         
\usepackage{hhline,multirow}  
\usepackage{dcolumn}          
\usepackage{slashed}
\usepackage{datetime}
\usepackage{color}
\usepackage[dvipsnames]{xcolor}
\usepackage{slashed}
\usepackage{float}

\usepackage{amscd}            
\usepackage{epsfig}
\usepackage{dcolumn}          
\usepackage[dvipsnames]{xcolor}
\usepackage[normalem]{ulem}
\usepackage{mathtools}
\usepackage{comment}
\usepackage[utf8]{inputenc}

\RequirePackage{color}
\RequirePackage[colorlinks=true
,urlcolor=black
,anchorcolor=black
,citecolor=black
,filecolor=black
,linkcolor=black
,menucolor=black
,pagecolor=black
,linktocpage=black
,pdfproducer=medialab
,pdfa=true
]{hyperref}

\newcommand{\nocontentsline}[3]{}
\newcommand{\tocless}[2]{\bgroup\let\addcontentsline=\nocontentsline#1{#2}\egroup}

\allowdisplaybreaks[2]

\begin{document}

\preprint{JLAB-THY-18-2757}
\preprint{NIKHEF 2018-032}

\title{Effect of flavor-dependent partonic transverse momentum \\ on the determination of the $W$ boson mass in hadronic collisions}

\newcommand*{\PaviaU}{Dipartimento di Fisica, Universit\`a di Pavia,
  via Bassi 6, I-27100 Pavia, Italy}\affiliation{\PaviaU}
\newcommand*{\InfnPavia}{INFN, Sezione di Pavia,
  via Bassi 6, I-27100 Pavia, Italy}\affiliation{\InfnPavia}
\newcommand*{\Nikhef}{Nikhef,
 Science Park 105, NL-1098 XG Amsterdam, the Netherlands}\affiliation{\Nikhef}
\newcommand*{\JLab}{Theory Center, Thomas Jefferson National Accelerator Facility \\ 12000 Jefferson Avenue, Newport News, VA 23606, USA}\affiliation{\JLab}


\author{Alessandro Bacchetta}
\thanks{ORCID: http://orcid.org/0000-0002-8824-8355} 
\affiliation{\PaviaU}\affiliation{\InfnPavia}

\author{Giuseppe Bozzi}
\thanks{ORCID: http://orcid.org/0000-0002-2908-6077}
\affiliation{\PaviaU}\affiliation{\InfnPavia}

\author{Marco Radici}
\thanks{ORCID: http://orcid.org/0000-0002-4542-9797}
\affiliation{\InfnPavia} 

\author{Mathias Ritzmann}
\thanks{mathias.ritzmann@gmail.com}
\affiliation{\Nikhef} 

\author{Andrea Signori}
\thanks{ORCID: http://orcid.org/0000-0001-6640-9659}
\affiliation{\JLab}

\date{\today}

\begin{abstract} 
Within the framework of transverse-momentum-dependent factorization, we investigate for the first time the impact of a flavor-dependent intrinsic transverse momentum of quarks on the production of $W^{\pm}$ bosons in proton-proton collisions at $\sqrt{s}$ = 7 TeV. We estimate the shift in the extracted value of the $W$ boson mass $M_W$ induced by different choices of flavor-dependent parameters for the intrinsic quark transverse momentum by means of a template fit to the transverse-mass and the lepton transverse-momentum distributions of the $W$-decay products. 
We obtain $-6\leq \Delta M_{W^+} \leq 9$ MeV and $-4\leq \Delta M_{W^-} \leq 3$ MeV with a statistical uncertainty of $\pm 2.5$ MeV. Our findings call for more detailed investigations of flavor-dependent nonperturbative effects linked to the proton structure at hadron colliders.
\end{abstract}

\maketitle

\textbf{\textit{Introduction and motivation.}}

Nonperturbative effects in transverse-momentum-dependent (TMD) phenomena are a central topic in the hadronic physics community with potentially important applications to high-energy physics.
The study of nonperturbative corrections originates from the work of Parisi and Petronzio~\cite{Parisi:1979se} and Collins, Soper, and Sterman~\cite{Collins:1984kg}, which focused on the role of the hard scale of the process compared to the infrared scale of QCD.
TMD factorization and evolution have been extensively studied in the literature~\cite{Collins:2011zzd,GarciaEchevarria:2011rb,Rogers:2015sqa,Collins:2017oxh}, together with the matching to collinear factorization~\cite{Collins:1984kg,Arnold:1990yk,Nadolsky:2002jr,Berger:2004cc,Stewart:2013faa,Collins:2016hqq,Echevarria:2018qyi}. 
Despite the limited amount of data available and the many open theoretical questions, in the past years we started gaining phenomenological information about TMD parton distribution functions (TMD PDFs) with increasing level of accuracy. Recently, the unpolarized quark TMD PDF was extracted for the first time from a global fit of data for semi-inclusive deep-inelastic scattering (SIDIS) and production of Drell-Yan lepton pairs and Z bosons~\cite{Bacchetta:2017gcc}. Nonetheless, we need a deeper understanding of many theoretical and phenomenological aspects~\cite{Angeles-Martinez:2015sea}.

In this paper, we demonstrate that if we want to determine the free parameters of the Standard Model with very high precision, then the effects of a possible flavor dependence of the intrinsic partonic transverse momentum should not be neglected even in the kinematic region where nonperturbative effects are expected to be small~\cite{Berger:2002ut,Berger:2003pd,Berger:2004cc,Echevarria:2012pw,DAlesio:2014mrz,Bacchetta:2017gcc,Scimemi:2017etj} ($\Lambda_{\mathrm{QCD}} \ll Q \ll \sqrt{s}$: $W$ boson production at the LHC lies in this kinematic region). In particular, we focus on the impact of the simplest TMD PDF, the unpolarized one, on the determination of the $W$ boson mass at hadron colliders.  \\

\textbf{\textit{Experimental measurements and uncertainties.}}

The determination of the $W$ boson mass, $M_{W}$, from the global electroweak fit ($M_{W}$=80.356$\pm$0.008 GeV)~\cite{Baak:2014ora} features a very small uncertainty that sets a goal for the precision of the experimental measurements at hadron colliders. 

Precise determinations of $M_{W}$ have been extracted from $p\bar{p}$ collisions at {\tt D0}~\cite{D0:2013jba} and at {\tt CDF}~\cite{Aaltonen:2013vwa}, and from $pp$ collisions at {\tt ATLAS}~\cite{Aaboud:2017svj} with a total uncertainty of 23 MeV, 19 MeV and 19 MeV, respectively. The current world average, based on these measurements and the ones performed at LEP, is $M_{W}$=80.379$\pm$0.012 GeV~\cite{PDG}. 
The experimental analyses are based on a template-fit procedure on the differential distributions of the decay products: in particular, the transverse momentum of the final lepton, $p_{T\ell}$, the transverse momentum of the neutrino $p_{T\nu}$ (only at the Tevatron), and the transverse mass $m_{T}$ of the lepton pair (where $m_T=\sqrt{2\;p_{T\ell}\;p_{T\nu}\;(1-\cos(\phi_{\ell}-\phi_{\nu}))}$, with $\phi_{\ell , \nu}$ being the azimuthal angles of the lepton and the neutrino, respectively). 

In a template-fit procedure, several histograms are generated with the highest available theoretical accuracy and the best available description of detector effects, letting the fit parameter ($M_{W}$, in this case) vary in a range: the histogram best describing experimental data selects the measured value for $M_{W}$. The details of the theoretical calculations used to compute the templates (choice of scales, PDFs, perturbative order, resummation of logarithmically enhanced contributions, nonperturbative effects, \dots) affect the result of the fit and define the theoretical systematics~\cite{CarloniCalame:2016ouw}. In this work we focus only on the impact of nonperturbative effects and, in particular, on those coming from the intrinsic transverse momentum of the initial-state partons. These effects modify the spectrum of the $W$ transverse momentum, $q_{T}^{W}$, subsequently inducing a nonnegligible shift in the extracted value of $M_{W}$. 


The three experimental collaborations {\tt D0, CDF}, and {\tt ATLAS} tipically fit the $Z$ data to obtain an estimate for the nonperturbative parameters. Then, assuming the parameters to be flavor independent, they use these estimates to predict the $q_T^W$ distribution. 
The uncertainty on $M_{W}$ due to the modelling of $q_{T}^{W}$ via template fits for the distributions in ($m_{T}, p_{T\ell}, p_{T\nu}$) are, respectively, $\delta M_{W}$ = (3,9,4) MeV for {\tt CDF}~\cite{Aaltonen:2013vwa}, $\delta M_{W}$ = (2,5,2) MeV for {\tt D0}~\cite{D0:2013jba} and $\delta M_{W}$ = (3,3) MeV for {\tt ATLAS}~\cite{Aaboud:2017svj} (the {\tt ATLAS} analysis did not include $p_{T\nu}$ in the template fit).

It is well known that one of the largest sources of error in determining $M_W$ comes from the uncertainty in the choice of the collinear PDFs~\cite{Bozzi:2011ww,Bozzi:2015hha,Bozzi:2015zja,Quackenbush:2015yra}. Nevertheless, one can see that the uncertainty propagating from the $q_T^W$ spectrum via $p_{T\ell}$ can be likewise comparably large (except for {\tt ATLAS}, because of the narrow range used for the $p_{T\ell}$ fit with respect to the $m_{T}$ one). This does not come as a surprise, since the $p_{T\ell}$ distribution is extremely sensitive to the modelling of $q_{T}^{W}$, i.e., the $p_{T\ell}$ shape gets more distorted by all-order resummation and nonperturbative contributions than the $m_{T}$ shape (which, in turn, is dominated by detector resolution). 

At present, neither analyses at the Tevatron and at the LHC included information on the flavor dependence of the intrinsic transverse momentum of the incoming partons participating in the hard scattering. Here, it is our aim to study its impact onto the determination of $M_W$ in hadronic collisions, taking inspiration from the phenomenological extraction of the unpolarized TMD PDF from low-energy data~\cite{Signori:2013mda}. \\

\textbf{\textit{Formalism.}}

The impact of nonperturbative effects in Drell-Yan and Higgs production has been extensively investigated (see, e.g.,~\cite{Kulesza:2003wi, Guzzi:2013aja, DAlesio:2014mrz, Scimemi:2017etj, Bacchetta:2017gcc,Chen:2018pzu,Bizon:2018foh,Alioli:2016fum} for available calculations and fitting codes). 

Different implementations of the nonperturbative contributions have been presented in the literature (see e.g. Refs.~\cite{Scimemi:2017etj, Bacchetta:2017gcc} and references therein).
In order to take into account possible differences between the valence and the sea quarks (and among different flavors in general), a flavor- and kinematic-dependent implementation of the nonperturbative part of the quark Sudakov exponent has been suggested in Refs.~\cite{Signori:2013mda, Bacchetta:2015ora}. 
In the present work, we choose a Gaussian functional form for the intrinsic transverse momentum distribution of the unpolarized TMD PDF. Its Fourier-conjugate expression reads 
\begin{equation}
f_1^{a NP}(b_T^2) \propto  \, e^{- g_{NP}^a  b_T^2}\ ,
\label{e:gauss_NP_TMDPDF}
\end{equation}
where $g_{NP}^a$ is related to the average intrinsic transverse momentum squared of a parton with flavor $a$. In general, the latter may also depend on kinematics, but here we will neglect this dependence. 



We implemented the above ansatz in two publicly available tools for computing Drell-Yan differential cross sections: {\tt DYqT}~\cite{Bozzi:2008bb, Bozzi:2010xn} and {\tt DYRes}~\cite{Bozzi:2008bb, Catani:2015vma}. 
The {\tt DYqT} program computes the $q_{T}$ spectrum of an electroweak boson $V$ ($V = \gamma^{*}, W^{\pm}, Z$) produced in hadronic collisions. The calculation combines the pure fixed-order QCD result up to ${\cal O}(\alpha^{2}_{s})$ at high $q_{T}$ ($q_{T}\sim M_{V}$) with the resummation of the logarithmically-enhanced contributions at small transverse-momenta ($q_{T}\ll M_{V}$) up to next-to-next-to-leading logarithmic (NNLL) accuracy. The rapidity of the vector boson and the leptonic kinematical variables are integrated over the entire kinematical range. 
At the same perturbative accuracy, the {\tt DYRes} code also provides the full kinematics of the vector boson and of its decay products. It thus allows for the application of arbitrary cuts on the final-state kinematical variables and gives differential distributions in form of bin histograms, directly comparable to experimental measurements. 

The original codes implement the nonperturbative TMD effects as a flavor- and kinematic-independent Gaussian exponential $e^{-g_{NP}b_{T}^{2}}$ whose strength is governed by a single parameter $g_{NP}$ tuned at the electroweak scale. 
This factor incorporates the nonperturbative effects from both the TMD PDFs entering the cross section, including their evolution. 
In order to mimic a flavor dependence in each partonic intrinsic transverse momentum, we modify this simple implementation by decomposing $g_{NP}$ into the sum $g_{NP}^a + g_{NP}^{a'}$, where the flavor indices span the range $a,a' =u_v, u_s, d_v, d_s, s, c, b, g$ (the subscripts referring to the valence and sea components, respectively). For each parton with flavor $a$, the nonperturbative contribution $f_1^{a NP}$ of Eq.~\eqref{e:gauss_NP_TMDPDF} is included in the corresponding term in the flavor sum of the TMD factorization formula~\cite{Collins:2011zzd}. In the following, we assume $g_{NP}^{s}=g_{NP}^{c}=g_{NP}^{b}=g_{NP}^{g}$, i.e., we assume that in total the intrinsic transverse-momentum depends on five flavors.\\

\textbf{\textit{Analysis strategy.}}

The phenomenological extraction of Ref.~\cite{Signori:2013mda} is based on  about 1500 data points, however the nonperturbative parameters $g_{NP}^a$ in Eq.~\eqref{e:gauss_NP_TMDPDF} are not tightly constrained. 
A fit to $Z/\gamma^{*}$ data from Tevatron produces the value $g_{NP}\sim 0.8$ GeV$^2$ for the universal nonperturbative 
factor~\cite{Guzzi:2013aja}. We recall that this value refers to the convolution of two TMD PDFs inside the cross section; hence, each parton should equally contribute with a nonperturbative width of $\approx 0.4$ GeV$^2$. 
When we introduce the corresponding parameter $g_{NP}^a$ for a single TMD PDF with flavor $a$,  we split it as follows:
\begin{equation}
\exp(-g_{NP}^a  b_{T}^{2}) \longrightarrow \exp[-[g_{evo}\ln(Q^{2}/Q_{0}^{2})+g_a ] \, b_{T}^{2}] \ , 
\label{e:flavGauss}
\end{equation}
where the first term in the right hand side is the nonperturbative correction due the TMD PDF evolution, which is flavor independent (but, in principle, different for quarks and gluons), and $g_a$ is the genuine flavor-dependent contribution. Information on $g_{evo}$ can be deduced from Ref.~\cite{Bacchetta:2017gcc}, where the TMD PDF was extracted from the global fit of SIDIS, Drell-Yan and $Z$-production data ($g_{evo}$ corresponds to $g_2/4$ in Ref.~\cite{Bacchetta:2017gcc}). At $Q = M_{W}$ and $Q_{0} = 1$ GeV, we have $g_{evo} \ln(Q^{2}/Q_{0}^{2}) \approx 0.3$ GeV$^2$.  In order to account for the uncertainties affecting the determination of $g_{evo}$, we choose to consider the interval $[0.2, 0.6]$ GeV$^2$ as a reasonable range and we vary $g_a$ in Eq.~\eqref{e:flavGauss} such that the $g_{NP}^a$ values fall into this range.

Thus, we generate random widths in the allowed range for the considered five flavors. We build 50 sets of flavor-dependent parameters together with a flavor-independent set where all the parameters are put equal to the central value of the variation range, $\overline{g_{NP}^a} = 0.4$ GeV$^2$. Our analysis is performed by first selecting ``Z-equivalent" sets, and then making a template fit, as detailed here below.\\

{\it Selection of ``Z-equivalent'' sets.} For proton-proton collisions at $\sqrt s = 7$ TeV, we generate pseudodata for the $q_T$ distribution of the $Z$ boson (22 bins similar to the {\tt ATLAS} ones~\cite{Aaboud:2017svj}) using the flavor-independent set in the {\tt DYqT} code at ${\cal O}(\alpha_{s})$ and NLL accuracy. We do the same for proton-antiproton collisions at $\sqrt s = 1.96$ TeV (72 bins similar to the {\tt CDF} ones~\cite{Aaltonen:2013vwa}). We assign to each of the $q_{T}$ bins an uncertainty equal to the experimental one. We compute the $q_T$ distribution in the same conditions also for each of the 50 flavor-dependent sets. We calculate the $\chi^2$ between each of these 50 distributions and the pseudodata generated by the flavor-independent set. We retain only those flavor-dependent sets that have a $\chi^2 < 80$ on the ``{\tt CDF}-like" bins ($\chi^2$/d.o.f. $<1.1$) and a $\chi^2 < 44$ on the ``{\tt ATLAS}-like" bins ($\chi^2$/d.o.f. $<2$). The first criterion selects 48 flavor-dependent sets out of 50; only 30 sets out of 50 match the second one, because the {\tt ATLAS} data have smaller (experimental) uncertainties. We keep those flavor-dependent sets that fullfil both criteria. When considering all the bins, these sets have a total $\chi^2 < 124$ on the pseudodata ($\chi^2$/d.o.f. $<1.3$). In practice, these selected flavor-dependent sets are equivalent to the flavor-independent one (with which the $Z$ pseudodata are generated) at approximately $2\sigma$ level. Not surprisingly, this result implies that the $Z$ boson data alone are not able to discriminate between flavor-independent and flavor-dependent sets of nonperturbative parameters. Data from flavor-sensitive processes are needed, in particular from SIDIS~\cite{Airapetian:2012ki,Aghasyan:2017ctw,Boer:2011fh,Accardi:2012qut}. \\

{\it Template fit.} Following the scheme introduced in~\cite{CarloniCalame:2003ux,Bozzi:2011ww}, we perform a template fit to estimate the impact of our ``Z-equivalent'' flavor-dependent sets on the determination of $M_{W}$. 
We use the {\tt DYRes} code at the same accuracy (NLL at small transverse momentum and ${\cal O} (\alpha_s)$ at large transverse momentum) and kinematics as before, using the MSTW2008 NLO PDF set ~\cite{Martin:2009iq}, setting central values for the renormalization, factorization and resummation scales $\mu_{R}=\mu_{F}=\mu_{res}=M_{W}$, and implementing {\tt ATLAS} acceptance cuts on the final-state leptons~\cite{Aaboud:2017svj}. In {\tt DYRes}, the singularity of the resummed form factor at very large values of $b_{T}$ ($b_{T}\gtrsim1/\Lambda_{\mathrm{QCD}}$) is avoided by the usual $b_{*}$ prescription~\cite{Collins:1984kg}. Similarly, the correct behavior at very low $b_T$ is enforced by modifying the argument of the logarithmic terms as in Refs.~\cite{Bozzi:2008bb, Catani:2015vma}. 
The form factor in Eq.~\eqref{e:flavGauss} is usually interpreted as the nonperturbative contribution to TMD resummation for $b_{T}\gtrsim 1/\Lambda_{\mathrm{QCD}}$. 
We generate templates with very high statistics (750 M events) for the $m_{T}$, $p_{T\ell}$ distributions\footnote{Our analysis is performed on 30 bins in the interval $[60,90]$ GeV for $m_{T}$ and on 20 bins in the interval $[30,50]$ GeV for $p_{T\ell}$.} with different $M_{W}$ masses in the range 80.370 GeV $\leq M_{W} \leq$ 80.400 GeV, using the flavor-independent set for the nonperturbative parameters. Then, for each ``Z-equivalent'' flavor-dependent set we generate pseudodata with lower statistics (135 M events) for the same leptonic observables with the fixed value $M_{W} = 80.385$ GeV
. Finally, for each pseudodata set we compute the $\chi^{2}$ of the various templates and we identify the template with minimum $\chi^2$ in order to establish how large is the shift in $M_{W}$ induced by a particular choice of flavor-dependent nonperturbative parameters. The statistical uncertainty of the template-fit procedure has been estimated by considering statistically equivalent those templates for which $\Delta\chi^{2}=(\chi^{2}-\chi^{2}_{min})\leq1$. Consequently, we quote an uncertainty of 2.5 MeV for each of the obtained $M_{W}$ shifts.\\

\textbf{\textit{Impact on the $M_W$ determination.}}

The outcome of our template fit is summarized in Tabs.~\ref{t:NPsets} and~\ref{t:mw_shifts} for 5 representative sets out of the 30 ``Z-equivalent'' sets. The former table lists the values of the $g_{NP}^a$ parameter in Eq.~\eqref{e:flavGauss} for each of the 5 considered flavors $a=u_v, d_v, u_s, d_s, s=c=b=g$. The latter table shows the corresponding shifts induced in $M_{W}$ when applying our analysis to the $m_{T}$, $p_{T\ell}$ distributions for the $W^{+}$ and the $W^{-}$ production at the LHC ($\sqrt s$ = 7 TeV). 

%
\begin{table}[htp]
\begin{center}
\begin{tabular}{|c|c|c|c|c|c|}
\hline
Set & $u_{v}$ & $d_{v}$ & $u_{s}$ & $d_{s}$ & $s$ \\
\hline
1 & 0.34 & 0.26 & 0.46 & 0.59 & 0.32 \\
2 & 0.34 & 0.46 & 0.56 & 0.32 & 0.51 \\
3 & 0.55 & 0.34 & 0.33 & 0.55 & 0.30 \\
4 & 0.53 & 0.49 & 0.37 & 0.22 & 0.52 \\
5 & 0.42 & 0.38 & 0.29 & 0.57 & 0.27 \\
\hline
\end{tabular}
\end{center}
\caption{Values of the $g_{NP}^a$ parameter in Eq.~\eqref{e:flavGauss} for the flavors $a=u_{v},d_{v},u_{s},d_{s},s=c=b=g$. Units are GeV$^2$.
} 
\label{t:NPsets}
\end{table}

\begin{table}[htp]
\small
 \centering
\begin{tabular}{|c|cc|cc|}
  \hline
  \multicolumn{1}{|c|}{}&\multicolumn{2}{|c|}{$\Delta M_{W^+}$}&\multicolumn{2}{|c|}{$\Delta M_{W^-}$} \\
  \hline
  \hline
Set  & $m_{T}$ & $p_{T\ell}$ & $m_{T}$ & $p_{T\ell}$\\
\hline
1 & 0 & -1 & -2 & 3 \\              
2 & 0 & -6 & -2 & 0 \\
3 & -1 & 9 & -2 & -4 \\
4 & 0 & 0 & -2 & -4 \\
5 & 0 & 4 & -1 & -3 \\
\hline
\end{tabular}
\caption{Shifts in $M_{W^\pm}$ (in MeV) induced by the corresponding sets of flavor-dependent intrinsic transverse momenta outlined in Tab.~\ref{t:NPsets} (Statistical uncertainty: 2.5 MeV).}
\label{t:mw_shifts}
\end{table} 
As expected, the shifts induced by the analysis performed on $p_{T\ell}$ are generally larger than for the $m_{T}$ case, since the latter is less sensitive to $q_{T}^{W}$-modelling effects.

For set 3, the shift induced on $M_{W^+}$ by the $p_{T\ell}$ analysis is 9 MeV, its size is particularly large if compared to the corresponding uncertainty quoted by {\tt ATLAS} (3 MeV). In general, taking also into account the statistical uncertainty of our analysis, the absolute value of the shifts induced when considering the $p_{T\ell}$ observable could exceed 10 MeV. For $M_{W^-}$ the shifts are less significant and fall within a 2-$\sigma$ interval around zero.

In the kinematic conditions under consideration, $W^+$ bosons are dominantly produced by a $u \bar{d}$ partonic process, with the $u$ coming from the valence region. As a consequence, we observe that sets characterized by a larger value of the combination $g_{NP}^{u_v} + g_{NP}^{d_s}$ (sets 3 and 5) lead to positive shifts in the value of $M_{W^+}$, while sets with a smaller value of $g_{NP}^{u_v} + g_{NP}^{d_s}$ (set 2) lead to negative shifts. For $W^-$ the situation is less clear, because the dominant partonic channel is $\bar{u}d$, with similar contributions from the valence and sea components of the $d$ quark. It seems that sets with smaller values of the sum of $g_{NP}^{u_s} + g_{NP}^{d_v}+ g_{NP}^{u_s} + g_{NP}^{d_s}$ (sets 3, 4, 5) lead to negative shifts in the value of $M_{W^-}$. Set 1 has a large value of the sum of $g_{NP}^{u_s} + g_{NP}^{d_v}+ g_{NP}^{u_s} + g_{NP}^{d_s}$ and leads to a positive shift in $M_{W^-}$. Set 2, however, violates the expectations based on these simple arguments.

Different flavor-dependent sets may induce artificial asymmetric shifts for $M_{W^{+}}$ and $M_{W^{-}}$ in the flavor-independent template fits. For instance, if $M_{W^{-}} > M_{W^{+}}$ (which corresponds to the {\tt ATLAS} findings~\cite{Aaboud:2017svj}) a template fit to the $p_{T\ell}$ observable based on sets 1 and 2 would lead to different shifts $\Delta M_{W^{-}} > \Delta M_{W^{+}}$ such that the difference between the two masses is enhanced. In this case, a fit with the corresponding flavor-dependent nonperturbative contributions would lead to a reduction of the mass gap. On the contrary, using sets 3-5 one would obtain the opposite result. \\ 

\textbf{\textit{Outlook and future developments.}}

In this work, we investigated the uncertainties on the determination of $M_W$ at the LHC induced by a possible flavor dependence of the partonic intrinsic transverse momentum. From these outcomes, we point out that a ``flavor-blind'' data analysis may not be a sufficiently accurate option, especially when a total uncertainty lower than 10 MeV is expected for $M_{W}$ at the LHC~\cite{ATL-PHYS-PUB-2014-015}. 

Future data from flavor-sensitive processes such as SIDIS (from the 12 GeV upgrade at Jefferson Lab~\cite{Dudek:2012vr}, from the {\tt COMPASS} collaboration~\cite{Gautheron:2010wva}, and from a future Electron-Ion Collider with both proton and deuteron beams~\cite{Boer:2011fh,Accardi:2012qut}) will shed new light on the flavor decomposition of the unpolarized TMD PDF. These low-energy SIDIS data involve also the study of the flavor dependence in the fragmentation function (the unpolarized TMD FF). Therefore, new data from semi-inclusive $e^+ e^-$ annihilation will also be needed for the flavor decomposition of the TMD FF~\cite{Bacchetta:2015ora}. 

All these data will improve our knowledge of the partonic structure of hadrons, and may help in reducing the uncertainties in precision measurements at high energies. \\

\textbf{\textit{Acknowledgments.}}

Discussions with Giancarlo Ferrera, Guido Montagna, Piet Mulders, Oreste Nicrosini, Fulvio Piccinini, Jianwei Qiu and Alessandro Vicini are gratefully acknowledged. We are also grateful to Jos Vermaseren and the Nikhef CT Department for the support with computing facilities in the initial stages of this project. 
AS acknowledges support from U.S. Department of Energy contract DE-AC05-06OR23177, under which Jefferson Science Associates, LLC, manages and operates Jefferson Lab. The work of AS has been funded partly also by the program of the Stichting voor Fundamenteel Onderzoek der Materie (FOM), which is financially supported by the Nederlandse Organisatie voor Wetenschappelijk Onderzoek (NWO). 
This work is supported by the European Research Council (ERC) under the European Union's Horizon 2020 research and innovation program (grant agreement No. 647981, 3DSPIN).\\

\bibliography{ref_Wmass_flav,ref_exps}

\begin{thebibliography}{48}
\expandafter\ifx\csname natexlab\endcsname\relax\def\natexlab#1{#1}\fi
\expandafter\ifx\csname bibnamefont\endcsname\relax
  \def\bibnamefont#1{#1}\fi
\expandafter\ifx\csname bibfnamefont\endcsname\relax
  \def\bibfnamefont#1{#1}\fi
\expandafter\ifx\csname citenamefont\endcsname\relax
  \def\citenamefont#1{#1}\fi
\expandafter\ifx\csname url\endcsname\relax
  \def\url#1{\texttt{#1}}\fi
\expandafter\ifx\csname urlprefix\endcsname\relax\def\urlprefix{URL }\fi
\providecommand{\bibinfo}[2]{#2}
\providecommand{\eprint}[2][]{\url{#2}}

\bibitem[{\citenamefont{Parisi and Petronzio}(1979)}]{Parisi:1979se}
\bibinfo{author}{\bibfnamefont{G.}~\bibnamefont{Parisi}} \bibnamefont{and}
  \bibinfo{author}{\bibfnamefont{R.}~\bibnamefont{Petronzio}},
  \bibinfo{journal}{Nucl. Phys.} \textbf{\bibinfo{volume}{B154}},
  \bibinfo{pages}{427} (\bibinfo{year}{1979}).

\bibitem[{\citenamefont{Collins et~al.}(1985)\citenamefont{Collins, Soper, and
  Sterman}}]{Collins:1984kg}
\bibinfo{author}{\bibfnamefont{J.~C.} \bibnamefont{Collins}},
  \bibinfo{author}{\bibfnamefont{D.~E.} \bibnamefont{Soper}}, \bibnamefont{and}
  \bibinfo{author}{\bibfnamefont{G.~F.} \bibnamefont{Sterman}},
  \bibinfo{journal}{Nucl. Phys.} \textbf{\bibinfo{volume}{B250}},
  \bibinfo{pages}{199} (\bibinfo{year}{1985}).

\bibitem[{\citenamefont{Collins}(2011)}]{Collins:2011zzd}
\bibinfo{author}{\bibfnamefont{J.}~\bibnamefont{Collins}},
  \emph{\bibinfo{title}{Foundations of Perturbative QCD}}, Cambridge Monographs
  on Particle Physics, Nuclear Physics and Cosmology
  (\bibinfo{publisher}{Cambridge University Press}, \bibinfo{year}{2011}).

\bibitem[{\citenamefont{Echevarria et~al.}(2012)\citenamefont{Echevarria,
  Idilbi, and Scimemi}}]{GarciaEchevarria:2011rb}
\bibinfo{author}{\bibfnamefont{M.~G.} \bibnamefont{Echevarria}},
  \bibinfo{author}{\bibfnamefont{A.}~\bibnamefont{Idilbi}}, \bibnamefont{and}
  \bibinfo{author}{\bibfnamefont{I.}~\bibnamefont{Scimemi}},
  \bibinfo{journal}{JHEP} \textbf{\bibinfo{volume}{1207}}, \bibinfo{pages}{002}
  (\bibinfo{year}{2012}), \eprint{1111.4996}.

\bibitem[{\citenamefont{Rogers}(2016)}]{Rogers:2015sqa}
\bibinfo{author}{\bibfnamefont{T.~C.} \bibnamefont{Rogers}},
  \bibinfo{journal}{Eur. Phys. J.} \textbf{\bibinfo{volume}{A52}},
  \bibinfo{pages}{153} (\bibinfo{year}{2016}), \eprint{1509.04766}.

\bibitem[{\citenamefont{Collins and Rogers}(2017)}]{Collins:2017oxh}
\bibinfo{author}{\bibfnamefont{J.}~\bibnamefont{Collins}} \bibnamefont{and}
  \bibinfo{author}{\bibfnamefont{T.~C.} \bibnamefont{Rogers}},
  \bibinfo{journal}{Phys. Rev.} \textbf{\bibinfo{volume}{D96}},
  \bibinfo{pages}{054011} (\bibinfo{year}{2017}), \eprint{1705.07167}.

\bibitem[{\citenamefont{Arnold and Kauffman}(1991)}]{Arnold:1990yk}
\bibinfo{author}{\bibfnamefont{P.~B.} \bibnamefont{Arnold}} \bibnamefont{and}
  \bibinfo{author}{\bibfnamefont{R.~P.} \bibnamefont{Kauffman}},
  \bibinfo{journal}{Nucl. Phys.} \textbf{\bibinfo{volume}{B349}},
  \bibinfo{pages}{381} (\bibinfo{year}{1991}).

\bibitem[{\citenamefont{Nadolsky et~al.}(2003)\citenamefont{Nadolsky,
  Kidonakis, Olness, and Yuan}}]{Nadolsky:2002jr}
\bibinfo{author}{\bibfnamefont{P.~M.} \bibnamefont{Nadolsky}},
  \bibinfo{author}{\bibfnamefont{N.}~\bibnamefont{Kidonakis}},
  \bibinfo{author}{\bibfnamefont{F.~I.} \bibnamefont{Olness}},
  \bibnamefont{and} \bibinfo{author}{\bibfnamefont{C.~P.} \bibnamefont{Yuan}},
  \bibinfo{journal}{Phys. Rev.} \textbf{\bibinfo{volume}{D67}},
  \bibinfo{pages}{074015} (\bibinfo{year}{2003}), \eprint{hep-ph/0210082}.

\bibitem[{\citenamefont{Berger et~al.}(2005)\citenamefont{Berger, Qiu, and
  Wang}}]{Berger:2004cc}
\bibinfo{author}{\bibfnamefont{E.~L.} \bibnamefont{Berger}},
  \bibinfo{author}{\bibfnamefont{J.-w.} \bibnamefont{Qiu}}, \bibnamefont{and}
  \bibinfo{author}{\bibfnamefont{Y.-l.} \bibnamefont{Wang}},
  \bibinfo{journal}{Phys. Rev.} \textbf{\bibinfo{volume}{D71}},
  \bibinfo{pages}{034007} (\bibinfo{year}{2005}), \eprint{hep-ph/0404158}.

\bibitem[{\citenamefont{Stewart et~al.}(2014)\citenamefont{Stewart, Tackmann,
  Walsh, and Zuberi}}]{Stewart:2013faa}
\bibinfo{author}{\bibfnamefont{I.~W.} \bibnamefont{Stewart}},
  \bibinfo{author}{\bibfnamefont{F.~J.} \bibnamefont{Tackmann}},
  \bibinfo{author}{\bibfnamefont{J.~R.} \bibnamefont{Walsh}}, \bibnamefont{and}
  \bibinfo{author}{\bibfnamefont{S.}~\bibnamefont{Zuberi}},
  \bibinfo{journal}{Phys. Rev.} \textbf{\bibinfo{volume}{D89}},
  \bibinfo{pages}{054001} (\bibinfo{year}{2014}), \eprint{1307.1808}.

\bibitem[{\citenamefont{Collins et~al.}(2016)\citenamefont{Collins, Gamberg,
  Prokudin, Rogers, Sato, and Wang}}]{Collins:2016hqq}
\bibinfo{author}{\bibfnamefont{J.}~\bibnamefont{Collins}},
  \bibinfo{author}{\bibfnamefont{L.}~\bibnamefont{Gamberg}},
  \bibinfo{author}{\bibfnamefont{A.}~\bibnamefont{Prokudin}},
  \bibinfo{author}{\bibfnamefont{T.~C.} \bibnamefont{Rogers}},
  \bibinfo{author}{\bibfnamefont{N.}~\bibnamefont{Sato}}, \bibnamefont{and}
  \bibinfo{author}{\bibfnamefont{B.}~\bibnamefont{Wang}},
  \bibinfo{journal}{Phys. Rev.} \textbf{\bibinfo{volume}{D94}},
  \bibinfo{pages}{034014} (\bibinfo{year}{2016}), \eprint{1605.00671}.

\bibitem[{\citenamefont{Echevarria et~al.}(2018)\citenamefont{Echevarria,
  Kasemets, Lansberg, Pisano, and Signori}}]{Echevarria:2018qyi}
\bibinfo{author}{\bibfnamefont{M.~G.} \bibnamefont{Echevarria}},
  \bibinfo{author}{\bibfnamefont{T.}~\bibnamefont{Kasemets}},
  \bibinfo{author}{\bibfnamefont{J.-P.} \bibnamefont{Lansberg}},
  \bibinfo{author}{\bibfnamefont{C.}~\bibnamefont{Pisano}}, \bibnamefont{and}
  \bibinfo{author}{\bibfnamefont{A.}~\bibnamefont{Signori}},
  \bibinfo{journal}{Phys. Lett.} \textbf{\bibinfo{volume}{B781}},
  \bibinfo{pages}{161} (\bibinfo{year}{2018}), \eprint{1801.01480}.

\bibitem[{\citenamefont{Bacchetta et~al.}(2017)\citenamefont{Bacchetta,
  Delcarro, Pisano, Radici, and Signori}}]{Bacchetta:2017gcc}
\bibinfo{author}{\bibfnamefont{A.}~\bibnamefont{Bacchetta}},
  \bibinfo{author}{\bibfnamefont{F.}~\bibnamefont{Delcarro}},
  \bibinfo{author}{\bibfnamefont{C.}~\bibnamefont{Pisano}},
  \bibinfo{author}{\bibfnamefont{M.}~\bibnamefont{Radici}}, \bibnamefont{and}
  \bibinfo{author}{\bibfnamefont{A.}~\bibnamefont{Signori}},
  \bibinfo{journal}{JHEP} \textbf{\bibinfo{volume}{06}}, \bibinfo{pages}{081}
  (\bibinfo{year}{2017}), \eprint{1703.10157}.

\bibitem[{\citenamefont{Angeles-Martinez
  et~al.}(2015)}]{Angeles-Martinez:2015sea}
\bibinfo{author}{\bibfnamefont{R.}~\bibnamefont{Angeles-Martinez}}
  \bibnamefont{et~al.}, \bibinfo{journal}{Acta Phys. Polon.}
  \textbf{\bibinfo{volume}{B46}}, \bibinfo{pages}{2501} (\bibinfo{year}{2015}),
  \eprint{1507.05267}.

\bibitem[{\citenamefont{Berger and Qiu}(2003{\natexlab{a}})}]{Berger:2002ut}
\bibinfo{author}{\bibfnamefont{E.~L.} \bibnamefont{Berger}} \bibnamefont{and}
  \bibinfo{author}{\bibfnamefont{J.-w.} \bibnamefont{Qiu}},
  \bibinfo{journal}{Phys. Rev.} \textbf{\bibinfo{volume}{D67}},
  \bibinfo{pages}{034026} (\bibinfo{year}{2003}{\natexlab{a}}),
  \eprint{hep-ph/0210135}.

\bibitem[{\citenamefont{Berger and Qiu}(2003{\natexlab{b}})}]{Berger:2003pd}
\bibinfo{author}{\bibfnamefont{E.~L.} \bibnamefont{Berger}} \bibnamefont{and}
  \bibinfo{author}{\bibfnamefont{J.-w.} \bibnamefont{Qiu}},
  \bibinfo{journal}{Phys. Rev. Lett.} \textbf{\bibinfo{volume}{91}},
  \bibinfo{pages}{222003} (\bibinfo{year}{2003}{\natexlab{b}}),
  \eprint{hep-ph/0304267}.

\bibitem[{\citenamefont{Echevarria et~al.}(2013)\citenamefont{Echevarria,
  Idilbi, Schafer, and Scimemi}}]{Echevarria:2012pw}
\bibinfo{author}{\bibfnamefont{M.~G.} \bibnamefont{Echevarria}},
  \bibinfo{author}{\bibfnamefont{A.}~\bibnamefont{Idilbi}},
  \bibinfo{author}{\bibfnamefont{A.}~\bibnamefont{Schafer}}, \bibnamefont{and}
  \bibinfo{author}{\bibfnamefont{I.}~\bibnamefont{Scimemi}},
  \bibinfo{journal}{Eur.Phys.J.} \textbf{\bibinfo{volume}{C73}},
  \bibinfo{pages}{2636} (\bibinfo{year}{2013}), \eprint{1208.1281}.

\bibitem[{\citenamefont{D'Alesio et~al.}(2014)\citenamefont{D'Alesio,
  Echevarria, Melis, and Scimemi}}]{DAlesio:2014mrz}
\bibinfo{author}{\bibfnamefont{U.}~\bibnamefont{D'Alesio}},
  \bibinfo{author}{\bibfnamefont{M.~G.} \bibnamefont{Echevarria}},
  \bibinfo{author}{\bibfnamefont{S.}~\bibnamefont{Melis}}, \bibnamefont{and}
  \bibinfo{author}{\bibfnamefont{I.}~\bibnamefont{Scimemi}},
  \bibinfo{journal}{JHEP} \textbf{\bibinfo{volume}{11}}, \bibinfo{pages}{098}
  (\bibinfo{year}{2014}), \eprint{1407.3311}.

\bibitem[{\citenamefont{Scimemi and Vladimirov}(2018)}]{Scimemi:2017etj}
\bibinfo{author}{\bibfnamefont{I.}~\bibnamefont{Scimemi}} \bibnamefont{and}
  \bibinfo{author}{\bibfnamefont{A.}~\bibnamefont{Vladimirov}},
  \bibinfo{journal}{Eur. Phys. J.} \textbf{\bibinfo{volume}{C78}},
  \bibinfo{pages}{89} (\bibinfo{year}{2018}), \eprint{1706.01473}.

\bibitem[{\citenamefont{Baak et~al.}(2014)\citenamefont{Baak, C\'uth, Haller,
  Hoecker, Kogler, Monig, Schott, and Stelzer}}]{Baak:2014ora}
\bibinfo{author}{\bibfnamefont{M.}~\bibnamefont{Baak}},
  \bibinfo{author}{\bibfnamefont{J.}~\bibnamefont{C\'uth}},
  \bibinfo{author}{\bibfnamefont{J.}~\bibnamefont{Haller}},
  \bibinfo{author}{\bibfnamefont{A.}~\bibnamefont{Hoecker}},
  \bibinfo{author}{\bibfnamefont{R.}~\bibnamefont{Kogler}},
  \bibinfo{author}{\bibfnamefont{K.}~\bibnamefont{Monig}},
  \bibinfo{author}{\bibfnamefont{M.}~\bibnamefont{Schott}}, \bibnamefont{and}
  \bibinfo{author}{\bibfnamefont{J.}~\bibnamefont{Stelzer}}
  (\bibinfo{collaboration}{Gfitter Group}), \bibinfo{journal}{Eur. Phys. J.}
  \textbf{\bibinfo{volume}{C74}}, \bibinfo{pages}{3046} (\bibinfo{year}{2014}),
  \eprint{1407.3792}.

\bibitem[{\citenamefont{Abazov et~al.}(2014)}]{D0:2013jba}
\bibinfo{author}{\bibfnamefont{V.~M.} \bibnamefont{Abazov}}
  \bibnamefont{et~al.} (\bibinfo{collaboration}{D0}), \bibinfo{journal}{Phys.
  Rev.} \textbf{\bibinfo{volume}{D89}}, \bibinfo{pages}{012005}
  (\bibinfo{year}{2014}), \eprint{1310.8628}.

\bibitem[{\citenamefont{Aaltonen et~al.}(2014)}]{Aaltonen:2013vwa}
\bibinfo{author}{\bibfnamefont{T.~A.} \bibnamefont{Aaltonen}}
  \bibnamefont{et~al.} (\bibinfo{collaboration}{CDF}), \bibinfo{journal}{Phys.
  Rev.} \textbf{\bibinfo{volume}{D89}}, \bibinfo{pages}{072003}
  (\bibinfo{year}{2014}), \eprint{1311.0894}.

\bibitem[{\citenamefont{Aaboud et~al.}(2018)}]{Aaboud:2017svj}
\bibinfo{author}{\bibfnamefont{M.}~\bibnamefont{Aaboud}} \bibnamefont{et~al.}
  (\bibinfo{collaboration}{ATLAS}), \bibinfo{journal}{Eur. Phys. J.}
  \textbf{\bibinfo{volume}{C78}}, \bibinfo{pages}{110} (\bibinfo{year}{2018}),
  \eprint{1701.07240}.

\bibitem[{\citenamefont{Tanabashi et~al.}(2018)}]{PDG}
\bibinfo{author}{\bibfnamefont{M.}~\bibnamefont{Tanabashi}}
  \bibnamefont{et~al.} (\bibinfo{collaboration}{Particle Data Group}),
  \bibinfo{journal}{Phys.Rev.} \textbf{\bibinfo{volume}{D98}},
  \bibinfo{pages}{030001} (\bibinfo{year}{2018}).

\bibitem[{\citenamefont{Carloni~Calame
  et~al.}(2017)\citenamefont{Carloni~Calame, Chiesa, Martinez, Montagna,
  Nicrosini, Piccinini, and Vicini}}]{CarloniCalame:2016ouw}
\bibinfo{author}{\bibfnamefont{C.~M.} \bibnamefont{Carloni~Calame}},
  \bibinfo{author}{\bibfnamefont{M.}~\bibnamefont{Chiesa}},
  \bibinfo{author}{\bibfnamefont{H.}~\bibnamefont{Martinez}},
  \bibinfo{author}{\bibfnamefont{G.}~\bibnamefont{Montagna}},
  \bibinfo{author}{\bibfnamefont{O.}~\bibnamefont{Nicrosini}},
  \bibinfo{author}{\bibfnamefont{F.}~\bibnamefont{Piccinini}},
  \bibnamefont{and} \bibinfo{author}{\bibfnamefont{A.}~\bibnamefont{Vicini}},
  \bibinfo{journal}{Phys. Rev.} \textbf{\bibinfo{volume}{D96}},
  \bibinfo{pages}{093005} (\bibinfo{year}{2017}), \eprint{1612.02841}.

\bibitem[{\citenamefont{Bozzi et~al.}(2011{\natexlab{a}})\citenamefont{Bozzi,
  Rojo, and Vicini}}]{Bozzi:2011ww}
\bibinfo{author}{\bibfnamefont{G.}~\bibnamefont{Bozzi}},
  \bibinfo{author}{\bibfnamefont{J.}~\bibnamefont{Rojo}}, \bibnamefont{and}
  \bibinfo{author}{\bibfnamefont{A.}~\bibnamefont{Vicini}},
  \bibinfo{journal}{Phys. Rev.} \textbf{\bibinfo{volume}{D83}},
  \bibinfo{pages}{113008} (\bibinfo{year}{2011}{\natexlab{a}}),
  \eprint{1104.2056}.

\bibitem[{\citenamefont{Bozzi et~al.}(2015{\natexlab{a}})\citenamefont{Bozzi,
  Citelli, and Vicini}}]{Bozzi:2015hha}
\bibinfo{author}{\bibfnamefont{G.}~\bibnamefont{Bozzi}},
  \bibinfo{author}{\bibfnamefont{L.}~\bibnamefont{Citelli}}, \bibnamefont{and}
  \bibinfo{author}{\bibfnamefont{A.}~\bibnamefont{Vicini}},
  \bibinfo{journal}{Phys. Rev.} \textbf{\bibinfo{volume}{D91}},
  \bibinfo{pages}{113005} (\bibinfo{year}{2015}{\natexlab{a}}),
  \eprint{1501.05587}.

\bibitem[{\citenamefont{Bozzi et~al.}(2015{\natexlab{b}})\citenamefont{Bozzi,
  Citelli, Vesterinen, and Vicini}}]{Bozzi:2015zja}
\bibinfo{author}{\bibfnamefont{G.}~\bibnamefont{Bozzi}},
  \bibinfo{author}{\bibfnamefont{L.}~\bibnamefont{Citelli}},
  \bibinfo{author}{\bibfnamefont{M.}~\bibnamefont{Vesterinen}},
  \bibnamefont{and} \bibinfo{author}{\bibfnamefont{A.}~\bibnamefont{Vicini}},
  \bibinfo{journal}{Eur. Phys. J.} \textbf{\bibinfo{volume}{C75}},
  \bibinfo{pages}{601} (\bibinfo{year}{2015}{\natexlab{b}}),
  \eprint{1508.06954}.

\bibitem[{\citenamefont{Quackenbush and Sullivan}(2015)}]{Quackenbush:2015yra}
\bibinfo{author}{\bibfnamefont{S.}~\bibnamefont{Quackenbush}} \bibnamefont{and}
  \bibinfo{author}{\bibfnamefont{Z.}~\bibnamefont{Sullivan}},
  \bibinfo{journal}{Phys. Rev.} \textbf{\bibinfo{volume}{D92}},
  \bibinfo{pages}{033008} (\bibinfo{year}{2015}), \eprint{1502.04671}.

\bibitem[{\citenamefont{Signori et~al.}(2013)\citenamefont{Signori, Bacchetta,
  Radici, and Schnell}}]{Signori:2013mda}
\bibinfo{author}{\bibfnamefont{A.}~\bibnamefont{Signori}},
  \bibinfo{author}{\bibfnamefont{A.}~\bibnamefont{Bacchetta}},
  \bibinfo{author}{\bibfnamefont{M.}~\bibnamefont{Radici}}, \bibnamefont{and}
  \bibinfo{author}{\bibfnamefont{G.}~\bibnamefont{Schnell}},
  \bibinfo{journal}{JHEP} \textbf{\bibinfo{volume}{1311}}, \bibinfo{pages}{194}
  (\bibinfo{year}{2013}), \eprint{1309.3507}.

\bibitem[{\citenamefont{Kulesza and Stirling}(2003)}]{Kulesza:2003wi}
\bibinfo{author}{\bibfnamefont{A.}~\bibnamefont{Kulesza}} \bibnamefont{and}
  \bibinfo{author}{\bibfnamefont{W.~J.} \bibnamefont{Stirling}},
  \bibinfo{journal}{JHEP} \textbf{\bibinfo{volume}{0312}}, \bibinfo{pages}{056}
  (\bibinfo{year}{2003}), \eprint{hep-ph/0307208}.

\bibitem[{\citenamefont{Guzzi et~al.}(2014)\citenamefont{Guzzi, Nadolsky, and
  Wang}}]{Guzzi:2013aja}
\bibinfo{author}{\bibfnamefont{M.}~\bibnamefont{Guzzi}},
  \bibinfo{author}{\bibfnamefont{P.~M.} \bibnamefont{Nadolsky}},
  \bibnamefont{and} \bibinfo{author}{\bibfnamefont{B.}~\bibnamefont{Wang}},
  \bibinfo{journal}{Phys.Rev.} \textbf{\bibinfo{volume}{D90}},
  \bibinfo{pages}{014030} (\bibinfo{year}{2014}), \eprint{1309.1393}.

\bibitem[{\citenamefont{Chen et~al.}(2018)\citenamefont{Chen, Gehrmann, Glover,
  Huss, Li, Neill, Schulze, Stewart, and Zhu}}]{Chen:2018pzu}
\bibinfo{author}{\bibfnamefont{X.}~\bibnamefont{Chen}},
  \bibinfo{author}{\bibfnamefont{T.}~\bibnamefont{Gehrmann}},
  \bibinfo{author}{\bibfnamefont{E.~W.~N.} \bibnamefont{Glover}},
  \bibinfo{author}{\bibfnamefont{A.}~\bibnamefont{Huss}},
  \bibinfo{author}{\bibfnamefont{Y.}~\bibnamefont{Li}},
  \bibinfo{author}{\bibfnamefont{D.}~\bibnamefont{Neill}},
  \bibinfo{author}{\bibfnamefont{M.}~\bibnamefont{Schulze}},
  \bibinfo{author}{\bibfnamefont{I.~W.} \bibnamefont{Stewart}},
  \bibnamefont{and} \bibinfo{author}{\bibfnamefont{H.~X.} \bibnamefont{Zhu}}
  (\bibinfo{year}{2018}), \eprint{1805.00736}.

\bibitem[{\citenamefont{Bizoń et~al.}(2018)\citenamefont{Bizoń, Chen,
  Gehrmann-De~Ridder, Gehrmann, Glover, Huss, Monni, Re, Rottoli, and
  Torrielli}}]{Bizon:2018foh}
\bibinfo{author}{\bibfnamefont{W.}~\bibnamefont{Bizoń}},
  \bibinfo{author}{\bibfnamefont{X.}~\bibnamefont{Chen}},
  \bibinfo{author}{\bibfnamefont{A.}~\bibnamefont{Gehrmann-De~Ridder}},
  \bibinfo{author}{\bibfnamefont{T.}~\bibnamefont{Gehrmann}},
  \bibinfo{author}{\bibfnamefont{N.}~\bibnamefont{Glover}},
  \bibinfo{author}{\bibfnamefont{A.}~\bibnamefont{Huss}},
  \bibinfo{author}{\bibfnamefont{P.~F.} \bibnamefont{Monni}},
  \bibinfo{author}{\bibfnamefont{E.}~\bibnamefont{Re}},
  \bibinfo{author}{\bibfnamefont{L.}~\bibnamefont{Rottoli}}, \bibnamefont{and}
  \bibinfo{author}{\bibfnamefont{P.}~\bibnamefont{Torrielli}}
  (\bibinfo{year}{2018}), \eprint{1805.05916}.

\bibitem[{\citenamefont{Alioli et~al.}(2017)}]{Alioli:2016fum}
\bibinfo{author}{\bibfnamefont{S.}~\bibnamefont{Alioli}} \bibnamefont{et~al.},
  \bibinfo{journal}{Eur. Phys. J.} \textbf{\bibinfo{volume}{C77}},
  \bibinfo{pages}{280} (\bibinfo{year}{2017}), \eprint{1606.02330}.

\bibitem[{\citenamefont{Bacchetta et~al.}(2015)\citenamefont{Bacchetta,
  Echevarria, Mulders, Radici, and Signori}}]{Bacchetta:2015ora}
\bibinfo{author}{\bibfnamefont{A.}~\bibnamefont{Bacchetta}},
  \bibinfo{author}{\bibfnamefont{M.~G.} \bibnamefont{Echevarria}},
  \bibinfo{author}{\bibfnamefont{P.~J.~G.} \bibnamefont{Mulders}},
  \bibinfo{author}{\bibfnamefont{M.}~\bibnamefont{Radici}}, \bibnamefont{and}
  \bibinfo{author}{\bibfnamefont{A.}~\bibnamefont{Signori}},
  \bibinfo{journal}{JHEP} \textbf{\bibinfo{volume}{11}}, \bibinfo{pages}{076}
  (\bibinfo{year}{2015}), \eprint{1508.00402}.

\bibitem[{\citenamefont{Bozzi et~al.}(2009)\citenamefont{Bozzi, Catani,
  Ferrera, de~Florian, and Grazzini}}]{Bozzi:2008bb}
\bibinfo{author}{\bibfnamefont{G.}~\bibnamefont{Bozzi}},
  \bibinfo{author}{\bibfnamefont{S.}~\bibnamefont{Catani}},
  \bibinfo{author}{\bibfnamefont{G.}~\bibnamefont{Ferrera}},
  \bibinfo{author}{\bibfnamefont{D.}~\bibnamefont{de~Florian}},
  \bibnamefont{and} \bibinfo{author}{\bibfnamefont{M.}~\bibnamefont{Grazzini}},
  \bibinfo{journal}{Nucl. Phys.} \textbf{\bibinfo{volume}{B815}},
  \bibinfo{pages}{174} (\bibinfo{year}{2009}), \eprint{0812.2862}.

\bibitem[{\citenamefont{Bozzi et~al.}(2011{\natexlab{b}})\citenamefont{Bozzi,
  Catani, Ferrera, de~Florian, and Grazzini}}]{Bozzi:2010xn}
\bibinfo{author}{\bibfnamefont{G.}~\bibnamefont{Bozzi}},
  \bibinfo{author}{\bibfnamefont{S.}~\bibnamefont{Catani}},
  \bibinfo{author}{\bibfnamefont{G.}~\bibnamefont{Ferrera}},
  \bibinfo{author}{\bibfnamefont{D.}~\bibnamefont{de~Florian}},
  \bibnamefont{and} \bibinfo{author}{\bibfnamefont{M.}~\bibnamefont{Grazzini}},
  \bibinfo{journal}{Phys. Lett.} \textbf{\bibinfo{volume}{B696}},
  \bibinfo{pages}{207} (\bibinfo{year}{2011}{\natexlab{b}}),
  \eprint{1007.2351}.

\bibitem[{\citenamefont{Catani et~al.}(2015)\citenamefont{Catani, de~Florian,
  Ferrera, and Grazzini}}]{Catani:2015vma}
\bibinfo{author}{\bibfnamefont{S.}~\bibnamefont{Catani}},
  \bibinfo{author}{\bibfnamefont{D.}~\bibnamefont{de~Florian}},
  \bibinfo{author}{\bibfnamefont{G.}~\bibnamefont{Ferrera}}, \bibnamefont{and}
  \bibinfo{author}{\bibfnamefont{M.}~\bibnamefont{Grazzini}},
  \bibinfo{journal}{JHEP} \textbf{\bibinfo{volume}{12}}, \bibinfo{pages}{047}
  (\bibinfo{year}{2015}), \eprint{1507.06937}.

\bibitem[{\citenamefont{Airapetian et~al.}(2013)}]{Airapetian:2012ki}
\bibinfo{author}{\bibfnamefont{A.}~\bibnamefont{Airapetian}}
  \bibnamefont{et~al.} (\bibinfo{collaboration}{HERMES}),
  \bibinfo{journal}{Phys. Rev.} \textbf{\bibinfo{volume}{D87}},
  \bibinfo{pages}{074029} (\bibinfo{year}{2013}), \eprint{1212.5407}.

\bibitem[{\citenamefont{Aghasyan et~al.}(2018)}]{Aghasyan:2017ctw}
\bibinfo{author}{\bibfnamefont{M.}~\bibnamefont{Aghasyan}} \bibnamefont{et~al.}
  (\bibinfo{collaboration}{COMPASS}), \bibinfo{journal}{Phys. Rev.}
  \textbf{\bibinfo{volume}{D97}}, \bibinfo{pages}{032006}
  (\bibinfo{year}{2018}), \eprint{1709.07374}.

\bibitem[{\citenamefont{Boer et~al.}(2011)\citenamefont{Boer, Diehl, Milner,
  Venugopalan, Vogelsang et~al.}}]{Boer:2011fh}
\bibinfo{author}{\bibfnamefont{D.}~\bibnamefont{Boer}},
  \bibinfo{author}{\bibfnamefont{M.}~\bibnamefont{Diehl}},
  \bibinfo{author}{\bibfnamefont{R.}~\bibnamefont{Milner}},
  \bibinfo{author}{\bibfnamefont{R.}~\bibnamefont{Venugopalan}},
  \bibinfo{author}{\bibfnamefont{W.}~\bibnamefont{Vogelsang}},
  \bibnamefont{et~al.} (\bibinfo{year}{2011}), \eprint{1108.1713}.

\bibitem[{\citenamefont{Accardi et~al.}(2016)}]{Accardi:2012qut}
\bibinfo{author}{\bibfnamefont{A.}~\bibnamefont{Accardi}} \bibnamefont{et~al.},
  \bibinfo{journal}{Eur. Phys. J.} \textbf{\bibinfo{volume}{A52}},
  \bibinfo{pages}{268} (\bibinfo{year}{2016}), \eprint{1212.1701}.

\bibitem[{\citenamefont{Carloni~Calame
  et~al.}(2004)\citenamefont{Carloni~Calame, Montagna, Nicrosini, and
  Treccani}}]{CarloniCalame:2003ux}
\bibinfo{author}{\bibfnamefont{C.~M.} \bibnamefont{Carloni~Calame}},
  \bibinfo{author}{\bibfnamefont{G.}~\bibnamefont{Montagna}},
  \bibinfo{author}{\bibfnamefont{O.}~\bibnamefont{Nicrosini}},
  \bibnamefont{and} \bibinfo{author}{\bibfnamefont{M.}~\bibnamefont{Treccani}},
  \bibinfo{journal}{Phys. Rev.} \textbf{\bibinfo{volume}{D69}},
  \bibinfo{pages}{037301} (\bibinfo{year}{2004}), \eprint{hep-ph/0303102}.

\bibitem[{\citenamefont{Martin et~al.}(2009)\citenamefont{Martin, Stirling,
  Thorne, and Watt}}]{Martin:2009iq}
\bibinfo{author}{\bibfnamefont{A.~D.} \bibnamefont{Martin}},
  \bibinfo{author}{\bibfnamefont{W.~J.} \bibnamefont{Stirling}},
  \bibinfo{author}{\bibfnamefont{R.~S.} \bibnamefont{Thorne}},
  \bibnamefont{and} \bibinfo{author}{\bibfnamefont{G.}~\bibnamefont{Watt}},
  \bibinfo{journal}{Eur. Phys. J.} \textbf{\bibinfo{volume}{C63}},
  \bibinfo{pages}{189} (\bibinfo{year}{2009}), \eprint{0901.0002}.

\bibitem[{ATL(2014)}]{ATL-PHYS-PUB-2014-015}
\bibinfo{type}{Tech. Rep.} \bibinfo{number}{ATL-PHYS-PUB-2014-015},
  \bibinfo{institution}{CERN}, \bibinfo{address}{Geneva}
  (\bibinfo{year}{2014}).

\bibitem[{\citenamefont{Dudek et~al.}(2012)}]{Dudek:2012vr}
\bibinfo{author}{\bibfnamefont{J.}~\bibnamefont{Dudek}} \bibnamefont{et~al.},
  \bibinfo{journal}{Eur. Phys. J.} \textbf{\bibinfo{volume}{A48}},
  \bibinfo{pages}{187} (\bibinfo{year}{2012}), \eprint{1208.1244}.

\bibitem[{\citenamefont{Gautheron et~al.}(2010)}]{Gautheron:2010wva}
\bibinfo{author}{\bibfnamefont{F.}~\bibnamefont{Gautheron}}
  \bibnamefont{et~al.} (\bibinfo{collaboration}{COMPASS}), \bibinfo{type}{Tech.
  Rep.} \bibinfo{number}{SPSC-P-340, CERN-SPSC-2010-014}
  (\bibinfo{year}{2010}).

\end{thebibliography}

\end{document}